\documentclass[11pt]{article}
\usepackage[margin=1in]{geometry}
\usepackage{amsmath,amssymb}
\usepackage{graphicx}
\usepackage{booktabs}
\usepackage{caption}
\usepackage{comment}
\usepackage{hyperref}
\usepackage[normalem]{ulem}
\hypersetup{
    colorlinks=true,
    linkcolor=black,
    citecolor=black,
    urlcolor=blue,
    pdftitle={Predicting Healthcare Provider Engagement in SMS Campaigns},
    pdfauthor={Daanish A. Qureshi, Rafay Chaudhary, Kok S. Tan, Or Maoz, Scott Burian, Michael Gelber, Phillip H. Kang, and Alan G. Labouseur}
}

\newlength{\bibitemsep}
\newlength{\bibparskip}
\let\oldthebibliography\thebibliography
\renewcommand\thebibliography[1]{%
  \oldthebibliography{#1}%
  \setlength{\parskip}{\bibitemsep}%
  \setlength{\itemsep}{\bibparskip}%
}

\title{Predicting Healthcare Provider Engagement in SMS Campaigns}

\author{
Daanish A. Qureshi\footnote{Corresponding author, Daanish.Qureshi@Impiricus.com}, Rafay Chaudhary, Kok S. Tan, Or Maoz, Scott Burian, \\Michael Gelber, Phillip H. Kang, and Alan G. Labouseur\\[6pt]
\small Impiricus, Atlanta, GA, USA
}

\date{}

\begin{document}
\maketitle

\begin{abstract}
As digital communication grows in importance when connecting with healthcare providers, traditional behavioral and content message features are imbued with renewed significance. If one is to meaningfully connect with them, it is crucial to understand what drives them to engage and respond. In this study, the authors analyzed several million text messages sent through the Impiricus platform to learn which factors influenced whether or not a doctor clicked on a link in a message. Several key insights came to light through the use of logistic regression, random forest, and neural network models, the details of which the authors discuss in this paper.

\end{abstract}


\section{Introduction}
Pharmaceutical companies have been educating healthcare providers (HCPs) about new medicines and treatments for decades, shaping patterns of care and influencing treatment decisions. 
Traditionally, these educational conversations happened in person.
But as hospitals and clinics have limited in-person visits in recent years, companies have increasingly turned to digital communication \cite{larkin2017association}. Today, pharmaceutical companies connect with HCPs using many online tools: e-mail, digital advertisements, virtual meetings, and even professional social media platforms \cite{hincapie2021perceptions,manz2014marketing,diekema2022health,ge2023digital,syrkiewicz2016perspectives,rahmanti2021social}.
And now, short message service (SMS) text messaging has emerged as a powerful digital tool. 
About 70\% of Americans report that texting is the fastest way to contact them, and nearly three-quarters read every text they receive \cite{Suffoletto2024}. 
In healthcare, SMS messaging has already been shown to drive deep behavioral changes. For example, it has been leveraged to help patients stay active, lose weight, and quit smoking \cite{orr2015mobile}.
Against this backdrop, the potential and interest in using SMS as a fast, reliable way to reach HCPs with professional information is clear.
In one study, HCPs who received medical reminders through text messages improved their knowledge and changed how they prescribed certain medications \cite{Chen2014}.
In another study, text messages were used to send quiz-style board review questions to medical residents who found them useful and relevant \cite{Nwigwe2023}. 
These examples reinforce the idea that SMS could play a key role in improving how doctors learn, stay informed, and adapt their clinical behavior. 
But how does one ensure that it actually works? \cite{gong2016leveraging}

Impiricus is a leading HCP engagement platform driving this new approach. Founded by physicians, the company enables healthcare organizations to connect directly with HCPs by delivering educational materials, clinical information, industry news, patient resources, and more straight to their mobile devices. By combining text messaging with other digital tools, Impiricus integrates SMS into a broader communication strategy to support marketing, education, and outreach initiatives. While early data show that these SMS campaigns can reach HCPs more effectively than traditional methods, there is still much to learn about what actually drives HCP engagement and which kinds of content are most effective.

The remainder of this paper is organized as follows: In Section~\ref{sec:Methods} we give background and present our dataset, pre-processing methods, statistical analyses, and machine-learning pipeline. 
In Section~\ref{sec:Results} we present our first contribution: the raw results of our study. 
Section~\ref{sec:Discussion} contains our second contribution: synthesizing our study results into a consistent hierarchy of engagement drivers.
Finally, Section~\ref{sec:Conclusions} concludes with what we have learned so far and suggests directions for future research.


\section{Methods}
\label{sec:Methods}

In this section we give some background context and present our dataset, pre-processing techniques, statistical analyses, and machine-learning pipeline. 

\subsection{Background}
One important way to measure engagement in text-based communication is through the click-through rate (CTR). CTR measures the percentage of recipients that click on a link in a message \cite{Reuter2021}. A higher CTR suggests that the message caught the reader’s attention and inspired them to take action. But the reasons why some doctors click while others do not
are not yet fully understood. 

Discovering patterns behind engagement can help improve how digital communications are designed and delivered, making SMS-based outreach to HCPs more effective, educational, and impactful.
To that end, we applied logistic regression, random forests, and neural network modeling techniques to millions of Impiricus text messages sent to healthcare professionals over two years. Analyzing this data, we were able to identify which factors predict whether or not a doctor will click on a message.

\subsection{Dataset}
We analyzed SMS messages sent to HCPs through the Impiricus platform in 2024 and 2025, a dataset containing millions of text message records. Each record represented a unique message-level interaction between the platform and an HCP, characterized by both message attributes and behavioral features derived from prior interactions.

Formally, each observation is represented as:
\[
x_i = [x_{i1}, x_{i2}, \dots, x_{in}], \quad y_i \in \{0,1\},
\]
where $x_i$ is the feature vector describing message and recipient characteristics, and $y_i$ is a binary indicator for whether the message was clicked ($y_i = 1$) or not ($y_i = 0$).


Key features include:
\begin{itemize}
    \item \textbf{National Provider Identifier (NPI)} — a unique provider identifier provided by the United States Department of Health and Human Services used for recipient-level linkage
    \item \textbf{Medical School Graduation Year} — proxy for experience (potentially reflecting generational engagement differences)
    \item \textbf{Self-reported and Claims-based Specialty} — a categorical feature encoding clinical area
    \item \textbf{Demeanor score ($d_i$)} — a sentiment-derived metric representing tone and valence of prior HCP responses
    \item \textbf{Message length ($\ell_i$)} — total number of characters in the outbound message
    \item \textbf{Links sent ($r_i$)} — number of hyperlinks included in the message
    \item \textbf{Click indicator ($y_i$)} — binary outcome indicating engagement
\end{itemize}

\subsection{Preprocessing Techniques}
Records missing key features or containing duplicate identifiers were excluded. Continuous variables were standardized to zero mean and unit variance:
\[
x' = \frac{x - \mu}{\sigma},
\]
where $\mu$ and $\sigma$ denote the feature mean and standard deviation, respectively.  
Categorical features (e.g., specialty) were one-hot encoded to create sparse binary representations.

Because most messages were not clicked, the dataset exhibited class imbalance. To avoid bias toward the majority (non-click) class, random undersampling was applied:
\[
P(y=1) \approx P(y=0),
\]
ensuring equal class representation during training and evaluation.

\subsection{Statistical Analyses}
Exploratory comparisons were performed to assess differences in message and behavioral features between clicked and non-clicked messages. For example, the mean message lengths for clicked ($\bar{\ell}_1$) and non-clicked ($\bar{\ell}_0$) groups were compared via a two-sample t-test:
\[
t = \frac{\bar{\ell}_1 - \bar{\ell}_0}{s_p \sqrt{\frac{1}{n_1} + \frac{1}{n_0}}},
\]
where $s_p$ is the pooled standard deviation and $n_1$, $n_0$ are group sample sizes.  
We also fit a logistic regression model to estimate directional feature effects:
\[
P(y_i = 1 \mid x_i) = \sigma(\beta_0 + \beta_1 x_{i1} + \cdots + \beta_p x_{ip}),
\]
where $\sigma(z) = \frac{1}{1 + e^{-z}}$ is the logistic function. Coefficients $\beta_j$ quantify the log-odds impact of each predictor on engagement likelihood.

\subsection{Machine-Learning Pipeline}
We trained three supervised models—logistic regression, random forest, and a feedforward neural network—implemented using Scikit-learn~\cite{pedregosa2011scikit} and PyTorch~\cite{paszke2019pytorch}.  
The dataset was split 80/20 into training and test sets, preserving balanced class proportions. Model hyperparameters were tuned via 5-fold cross-validation.

\paragraph{Logistic Regression.}  
Logistic regression served as the interpretable baseline model, optimized via maximum likelihood estimation:
\[
\hat{\beta} = \arg\max_\beta \sum_{i=1}^n \big[ y_i \log \sigma(x_i^\top \beta) + (1 - y_i) \log (1 - \sigma(x_i^\top \beta)) \big].
\]
This model provides directional coefficients and odds ratios, facilitating explainable feature-level interpretation.

\paragraph{Random Forest.}  
The random forest classifier aggregates an ensemble of $T$ decision trees $\{h_t(x)\}_{t=1}^T$, each trained on bootstrap samples with random feature subsets.  
Predictions are combined by majority vote:
\[
\hat{y} = \text{mode}\{h_1(x), h_2(x), \dots, h_T(x)\}.
\]
Feature importance was computed from the mean decrease in Gini impurity across all trees, reflecting each variable’s contribution to overall model performance.

\paragraph{Neural Network.}  
The neural network modeled nonlinear dependencies among behavioral and content features.  
It consisted of two fully connected hidden layers (128 and 64 neurons) with rectified linear unit (ReLU) activation:
\[
f(x) = \max(0, x).
\]
Dropout regularization (rate = 0.3) was applied between layers to prevent overfitting.  
The final layer used a sigmoid activation to predict engagement probability:
\[
\hat{y}_i = \sigma(W_2 f(W_1 x_i + b_1) + b_2).
\]
The model minimized binary cross-entropy loss:
\[
\mathcal{L} = -\frac{1}{n}\sum_{i=1}^n [y_i \log \hat{y}_i + (1 - y_i) \log (1 - \hat{y}_i)],
\]
using the Adam optimizer. Training and validation loss curves were monitored to ensure stable convergence without overfitting.

\paragraph{Evaluation Metrics.}  
Model performance was assessed on the test set using accuracy, precision, recall, F1-score, and area under the receiver operating characteristic curve (AUC).  
These metrics jointly measure both predictive accuracy and sensitivity to minority-class (clicked) cases.


\section{Results}
\label{sec:Results}

In this section we present our first contribution: the raw results of our study in terms of descriptive insights vis-a-vis logistic regression, random forest, and neural network models.

\subsection{Descriptive Insights}
After data cleaning and balancing, the final analytic dataset included 25{,}494 message–HCP interactions with equal representation of clicked and non-clicked outcomes. Clicked messages were substantially shorter on average than non-clicked messages (two-sample t-test: $T = 90.7344$, $p < 0.0001$). Engagement was also higher among HCPs with elevated demeanor scores, indicating that prior positive behavioral history strongly predicts future responsiveness. Specialty-based and demographic variables showed limited variation across groups, suggesting that behavioral and message-level factors drive the majority of predictive signal in click-through behavior.

\begin{table}[t]
\centering
\caption{Logistic Regression performance}
\label{tab:logreg}
\begin{tabular}{@{}lcccc@{}}
\toprule
Metric & Class 0 & Class 1 & Macro Avg & Weighted Avg \\
\midrule
Precision & 0.91 & 0.92 & 0.92 & 0.92 \\
Recall    & 0.93 & 0.91 & 0.92 & 0.92 \\
F1-score  & 0.92 & 0.91 & 0.92 & 0.92 \\
\hline
Accuracy  & \multicolumn{4}{c}{0.92} \\
AUC       & \multicolumn{4}{c}{0.94} \\
\bottomrule
\end{tabular}
\end{table}

\subsection{Logistic Regression}
The logistic regression model achieved an accuracy of 0.92 and an AUC of 0.94 (Table~\ref{tab:logreg}). Despite its linear form, the model revealed clear and interpretable directional effects. Demeanor exhibited a positive association with engagement ($\beta=0.0468$, OR $=1.05$), indicating that more positive tone in prior messages predicts higher responsiveness. In contrast, both message length ($\beta=-0.0063$, OR $=0.99$) and number of links ($\beta=-0.48$, OR $=0.62$) showed negative effects, implying that longer or link-heavy messages are less effective. 
Certain specialty categories demonstrated larger coefficients relative to the reference group—such as Pain Management (OR $=5.48$), Endocrinology (OR $=5.21$), and Ophthalmology (OR $=4.57$)—while Oncology/Hematology and Pediatric Oncology were associated with reduced engagement (ORs $<0.4$). These findings suggest that although most specialties exert minimal influence, a few subgroups may differ in baseline engagement behavior.

\subsection{Random Forest}
\begin{figure}[t]
    \centering
    \includegraphics[width=0.75\linewidth]{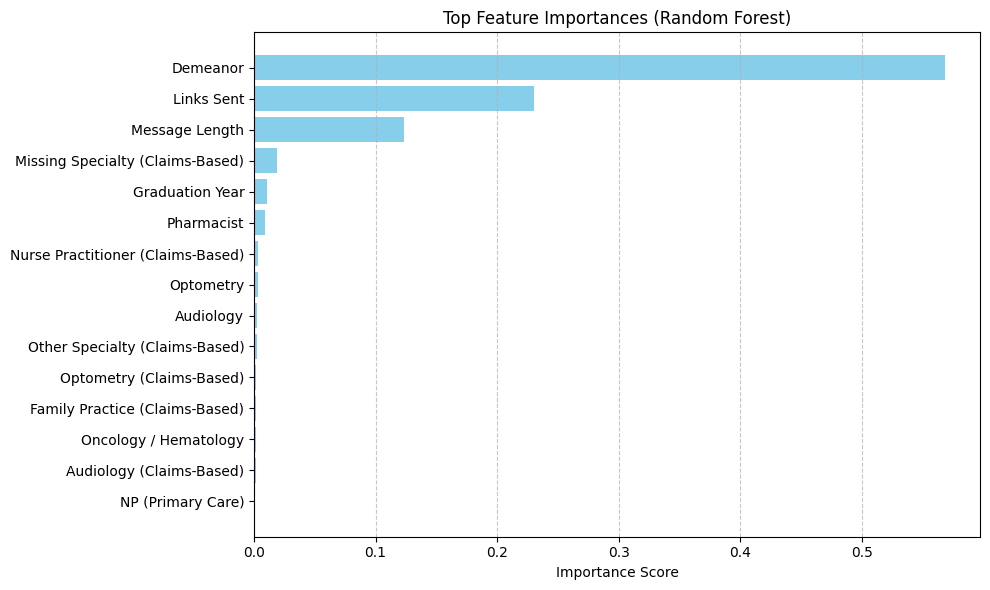}
    \caption{Top feature importances from the Random Forest model. Demeanor dominates, followed by links sent and message length.}
    \label{fig:fi}
\end{figure}

The random forest classifier improved upon the logistic baseline, achieving an accuracy of 0.92 and an AUC of 0.97 (Table~\ref{tab:rf}). Feature importance analysis (Figure~\ref{fig:fi}) confirmed that behavioral and content features dominate: demeanor emerged as the single strongest predictor, followed by links sent and message length. Specialty-based features contributed minimally, consistent with their limited explanatory power in the logistic model. 
The random forest captured nonlinear interactions—such as the compounding effect of positive demeanor and message brevity—while preserving directional agreement with the logistic regression’s coefficients. Importantly, the forest’s importance hierarchy parallels the odds ratio magnitudes from the logistic model, reinforcing that engagement depends primarily on behavioral tone and message structure rather than demographic or specialty characteristics.

\begin{table}[b]
\centering
\caption{Random Forest performance}
\label{tab:rf}
\begin{tabular}{@{}lcccc@{}}
\toprule
Metric & Class 0 & Class 1 & Macro Avg & Weighted Avg \\
\midrule
Precision & 0.91 & 0.93 & 0.92 & 0.92 \\
Recall    & 0.93 & 0.91 & 0.92 & 0.92 \\
F1-score  & 0.92 & 0.92 & 0.92 & 0.92 \\
\hline
Accuracy  & \multicolumn{4}{c}{0.92} \\
AUC       & \multicolumn{4}{c}{0.97} \\
\bottomrule
\end{tabular}
\end{table}

\subsection{Neural Network}
\begin{figure}[!b]
    \centering
    \includegraphics[width=0.75\linewidth]{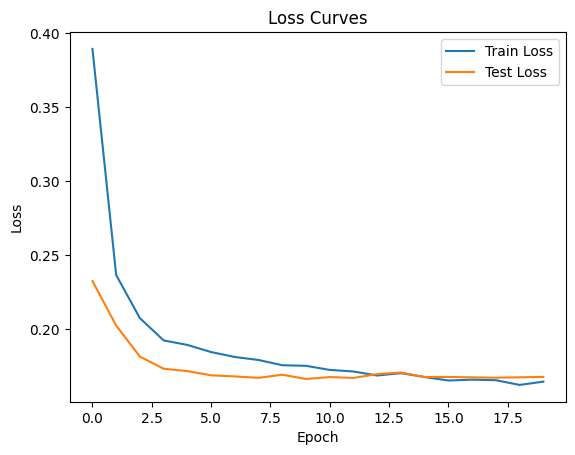}
    \caption{Training and validation loss curves for the Neural Network across 20 epochs, showing stable convergence without overfitting.}
    \label{fig:loss}
\end{figure}

The feedforward neural network achieved similarly strong performance, with an accuracy of 0.92 and an AUC of 0.96 (Table~\ref{tab:nn}). Training and validation loss curves (Figure~\ref{fig:loss}) demonstrated stable convergence and no evidence of overfitting. The model’s performance gain over logistic regression reflects its ability to capture subtle, nonlinear dependencies among demeanor, message length, and link count. However, interpretability remains limited relative to the other models. Overall, all three approaches converged on the same conclusion: that engagement is best predicted by concise, credible messages sent to HCPs with a history of positive behavioral interaction.

\begin{table}[h]
\centering
\caption{Neural Network performance.}
\label{tab:nn}
\begin{tabular}{@{}lcccc@{}}
\toprule
Metric & Class 0 & Class 1 & Macro Avg & Weighted Avg \\
\midrule
Precision & 0.91 & 0.93 & 0.92 & 0.92 \\
Recall    & 0.92 & 0.93 & 0.92 & 0.92 \\
F1-score  & 0.92 & 0.93 & 0.92 & 0.92 \\
\hline
Accuracy  & \multicolumn{4}{c}{0.92} \\
AUC       & \multicolumn{4}{c}{0.96} \\
\bottomrule
\end{tabular}
\end{table}

\section{Discussion}
\label{sec:Discussion}

In this section we synthesize our study results into our second contribution: a consistent hierarchy of engagement drivers.

Our results reinforce the idea that HCP engagement in SMS campaigns is driven primarily by prior behavioral and content-based features rather than static demographics. Across all models, the strongest predictors of click-through were demeanor (the tone of prior replies), the number of links sent, and message length. Shorter, positively toned messages consistently produced higher engagement rates, aligning with behavioral communication theory and the practical realities of clinicians’ limited attention.

The logistic regression model (Table~\ref{tab:logreg}) provided interpretable, directional insights into these relationships. Demeanor exhibited a positive association with engagement ($\beta=0.0468$, OR $=1.05$), indicating that HCPs with a history of courteous or positive replies are more likely to engage again. In contrast, both message length ($\beta=-0.0063$, OR $=0.99$) and number of links ($\beta=-0.48$, OR $=0.62$) showed negative effects, confirming that longer or link-heavy messages reduce the likelihood of clicks. A few specialty categories—such as Pain Management (OR $=5.48$), Endocrinology (OR $=5.21$), and Ophthalmology (OR $=4.57$)—showed localized positive effects relative to a reference group, while Oncology/Hematology and Pediatric Oncology were associated with lower engagement. However, specialty-level effects were small overall, suggesting that behavioral and content variables dominate predictive performance.

The random forest model complements these results by capturing nonlinear feature interactions while maintaining interpretability through feature importance rankings. Its hierarchy closely mirrored the logistic regression results: demeanor was the most influential predictor, followed by links sent and message length (Figure~\ref{fig:fi}). This alignment demonstrates model coherence—features that increase odds in the logistic model also drive the largest impurity reductions in the nonlinear model. The random forest thus substantiates the logistic regression findings, offering greater flexibility in modeling complex dependencies while confirming the same behavioral and content-based determinants of engagement.

The neural network achieved similarly high performance (Accuracy = 0.92, AUC = 0.96) but offered limited interpretability. Its small but consistent gain in AUC suggests that subtle nonlinear dependencies exist between tone, message structure, and perhaps certain specialties. However, because logistic regression and random forest already capture the dominant behavioral effects, the neural network’s improvements appear incremental rather than transformative. In operational settings—particularly within regulated healthcare communication contexts—model transparency and interpretability may outweigh marginal performance gains.

Taken together, all three models converge on a consistent hierarchy of engagement drivers: (1) positive historical behavioral tone, (2) fewer embedded links, and (3) concise message length. Logistic regression quantifies these relationships through interpretable odds ratios, the random forest confirms them as dominant global predictors under nonlinear conditions, and the neural network validates that higher-order interactions contribute modestly. These results collectively support a practical conclusion: to maximize HCP engagement, campaigns should prioritize brief, positively framed messages with minimal links, leveraging behavioral history as the most reliable indicator of responsiveness.


\section{Conclusions}
\label{sec:Conclusions}


This study identified key factors predicting HCP engagement in SMS campaigns. Three categories of features --- behavioral history, message structure, and limited specialty effects --- consistently explain variation in response. Messages sent to HCPs with a positive history of interaction (high demeanor), that are concise, and that contain fewer links are likely to generate engagement. These findings emphasize that tone, simplicity, and brevity together determine message effectiveness.

Across models, demeanor emerged as the most influential factor. Logistic regression quantified this relationship,
indicating that each unit increase in positive tone slightly raises the odds of engagement. In contrast, both message length and links sent were negatively associated with clicks, confirming that shorter and less link-dense messages perform better. The random forest validated this same hierarchy of predictors under nonlinear conditions, while the neural network achieved comparable performance
by capturing subtler interactions among behavioral and content-based variables. This coherence across models strengthens the evidence that behavioral signals and message design dominate over demographic or specialty-based features in predicting engagement.
Put simply: shorter and simpler messages with minimal links should be prioritized for outreach. Reducing cognitive and visual load while maintaining a positive and professional tone yields measurable improvements in response. 

Future extensions of this work could integrate temporal variables (e.g., time of day and day of week) and natural language features derived directly from message text. Reinforcement learning frameworks might then be applied to optimize message timing, tone, and structure in real time. 

Overall, these results demonstrate how predictive modeling can transform healthcare communication into a data-driven, personalized, and adaptive engagement strategy.

\bibliographystyle{unsrt}
\bibliography{references}

\end{document}